\documentclass[letterpaper,12pt]{article}
\usepackage{jheppub2} 
\usepackage[T1]{fontenc}

\usepackage{natbib}
\usepackage{graphicx}
\usepackage{amsmath}
\usepackage{amsfonts}
\usepackage{amssymb}
\usepackage{enumitem}
\usepackage{tocvsec2}

\newcommand\beq{\begin{eqnarray}}
\newcommand\eeq{\end{eqnarray}}

\hypersetup{
    colorlinks=true,   
    linkcolor=red,     
    citecolor=blue,    
    filecolor=magenta, 
    urlcolor=blue      
}

\makeatletter
\def\l@subsubsection#1#2{}
\makeatother    

\advance\footnotesep 1.5pt

\begin{document}

\title{Chern insulator transitions with Wilson fermions on a hyperrectangular lattice}

\author{Srimoyee Sen,}
\emailAdd{srimoyee08@gmail.com}
\affiliation{Department of Physics and Astronomy,  Iowa State University, Ames IA 50011, USA}

\abstract{
A $U(1)$ gauge theory coupled to a Wilson fermion on a $2+1$ dimensional cubic lattice is known to exhibit Chern insulator like topological transitions as a function of the the ratio $M/R$ where $M$ is the fermion mass and $R$ is the Wilson parameter.
 I show that, with $M$ and $R$ held fixed, a rectangular lattice with anisotropic lattice spacing can exhibit distinct topological phases as a function of the lattice anisotropy. 
As a consequence, a $2+1$ dimensional lattice theory without any domain wall in the fermion mass can still exhibit chiral edge modes on a $1+1$ dimensional defect across which lattice spacing changes abruptly. Likewise, a domain wall in the fermion mass on a uniform rectangular lattice can exhibit discrete changes in the number and chirality of zero modes as a function of lattice anisotropy. The construction presented in this paper can be generalized to higher dimensional space-time lattices.  
} 

\maketitle

\maxtocdepth{subsection} 
 
\section{Introduction}
The domain wall construction of chiral fermions in relativistic quantum field theories(QFT)
\cite{Callan:1984sa, Kaplan:1992bt} has close parallels with the physics of quantum Hall effect (QHE)\cite{PhysRevLett.45.494, PhysRevLett.49.405,PhysRevLett.54.259, PhysRevB.31.3372, PhysRevLett.61.2015}. Although quantum Hall states are realized in $2+1$ dimensions, the underlying principle of anomaly inflow and the associated edge modes generalize to higher dimensions in their relativistic QFT analog. For example, chiral domain wall fermions which are close cousins of the edge states in QHE can be realized on domain walls of space-time dimension $2n$ for any $n\geq 1$ with $n\in \mathcal{Z}$ embedded in one higher spatial dimension. 
The continuum construction of chiral fermionic zero modes and the corresponding anomaly inflow mechanism was first sketched out in a paper by Callan and Harvey\cite{Callan:1984sa}. The $2+1$ dimensional version of this construction involves a fermion with a spatially varying mass of the form $m \epsilon(x_{2})$ with $\epsilon(x_{2}>0)=1$ and $\epsilon(x_{2}<0)=-1$ coupled to a $U(1)$ gauge field. In the Infrared this theory exhibits a chiral zero mode localized on the domain wall at $x_2=0$. The bulk on the other hand exhibits a low energy Chern-Simons theory of level $1$ on one side of the wall and level $0$ on the other. This mimics the physics of QHE, anomalous quantum Hall effect to be precise\cite{doi:10.1146/annurev-conmatphys-031115-011417}, where the QHE sample is described by a nontrivial Chern-Simons theory and the edges of the sample exhibit chiral zero modes. \footnote{Other examples of topological phases and edge excitations in condensed matter systems that have analogs in relativistic QFTs include the quantum spin Hall effect \cite{Konig766,hsieh2008topological,Jansen:1992yj}, fractional quantum Hall effect \cite{Kaplan:2019pdd}, fractional quantum spin Hall effect \cite{Kaplan:2019pdd, PhysRevB.86.115131}, Majorana edge states \cite{Kaplan:1999jn, Endres:2009yp, Kitaev:2001kla}.} 

The continuum construction was subsequently adopted in lattice gauge theory \cite{Kaplan:1992bt} to circumvent the challenges of formulating chiral fermions on lattice \cite{Nielsen:1981hk}. This in turn revealed that lattice regularization can alter the infrared(IR) physics in ways that are inaccessible in the continuum. For example, the low energy effective field theory (EFT) of a massive fermion coupled to a $U(1)$ gauge field in $2+1$ dimensions is a level one Chern-Simons theory in the continuum. Lattice regularization modifies this EFT by introducing doublers which sets the Chern-Simons level to zero. In order to obtain a nontrivial Chern-Simons theory one needs to introduce a Wilson term for the fermions. With Wilson fermions of mass $M$ and Wilson parameter $R$, appropriately normalized in lattice units, the Chern-Simons level can be made to jump between $0, 1, -2, 1$ and $0$ as a function of $M/R$\cite{Golterman:1992ub}. These jumps are accompanied by commensurate changes in the number and chirality of zero modes on the domain wall \cite{Jansen:1992tw} so as to satisfy the anomaly inflow condition. The changes in the Chern-Simons level indicate that the bulk undergoes topological phase transitions as a function of $M/R$. These topological phases found in lattice gauge theory are in fact relativistic generalizations of the TKNN calculation\cite{PhysRevLett.49.405} performed in the context of QHE in a periodic potential and also have analogues in Dirac-Chern insulators\cite{PhysRevB.74.085308}.

The aforementioned calculation of topological phases in lattice gauge theory and the corresponding zero mode analysis were carried out on a cubic lattice \cite{Kaplan:1992bt, Golterman:1992ub, Jansen:1992yj} with the Wilson fermion mass and inverse Wilson parameter normalized in lattice units, i.e. the topological phase transitions take place at values $\frac{M}{R}=0,2,4$ and $6$ with $M=m a_l$ and $R=\frac{r}{a_l}$ where $m$ is the fermion mass, $r$ is the Wilson parameter and $a_l$ is the lattice spacing. Note that this definition of the Wilson parameter is slightly different from the standard convention where the dimensionless parameter $R$ is called the Wilson parameter. The continuum limit of this analysis is arrived at by taking $a_l \rightarrow 0$ while holding $\frac{M}{R}=\frac{m a_l^2}{r}$ fixed. Note that the ratio of parameters $\frac{m a_l^2}{r}$ dictating the topological behavior of the lattice theory involves lattice spacing $a_l$. This naturally raises the question :

{\it what is the phase diagram of a $U(1)$ gauge theory coupled to a Wilson fermion on a rectangular lattice (i.e. anisotropic lattice spacing) ?}

The goal of this paper is to address this question in the context of $2+1$ dimensional lattices\footnote{Anisotropic lattices have been explored in lattice gauge theory in the context of heavy quark \cite{Alford:1996nx, Klassen:1998ua}, hadron and glueball spectrum, scattering processes etc \cite{Burgio:2003in, PhysRevD.60.034509, Li:2007ey, PhysRevD.64.034509, PhysRevD.65.094508, PhysRevD.64.094509, Meng:2003gm, Du:2004ib}. Two nucleon systems were investigated on an anisotropic lattice in \cite{Detmold:2004qn}. Some of the earlier work on formulating $SU(N)$ gauge theories on anisotropic lattices can be found in \cite{Burgers:1987mb}, \cite{Karsch:1982ve}. Domain wall fermions were explored on anisotropic lattice in \cite{Feng:2006vt} where the physical four dimensional lattice was rectangular. Anisotropic Wilson gauge action was explored in \cite{Klassen:1998ua} and effective field theory of anisotropic Wilson lattice action was analyzed in \cite{Bedaque:2007xg}. However these studies do not explore the phase diagram of the lattice theory which is of interest to this paper.}. However, it is important to note that most of the analysis presented in this paper can be generalized to higher space-time dimensions. Also, the gauge sector of the theory can be modified to include $SU(N)$ gauge theories.

The anisotropic lattices considered in this paper consist of equal lattice spacing in the time ($x^0$ direction) and one of the two spatial dimensions ($x^1$ direction) whereas the lattice spacing in the remaining one spatial dimension $x_2$ is different. I call these two lattice spacings $a$ and $a_s$ respectively and define a lattice anisotropy parameter $\tilde{a}\equiv\frac{a}{a_s}$. The low energy Chern-Simons (C.S.) theory in the bulk after integrating out the Wilson fermions is found to depend on the ratio of the two lattice spacings. This result leads me to analyze two different types of defects
\begin{enumerate}
\item A domain wall in the fermion mass in $x^2$ direction on a uniform rectangular lattice, i.e. $a\neq a_s$.
\item An abrupt jump in the lattice spacing in the $x_2$ direction, i.e. $a_s(x_2>0)\neq a_s(x_2<0)$, while the Wilson fermion mass is spatially uniform. 
\end{enumerate} 
These defects which are $1+1$ dimensional walls transverse to the $x_2$ direction are sometimes called the physical lattice \cite{Feng:2006vt} and the choice of lattice spacings considered in this paper keeps these physical lattices cubic. For a domain wall in the fermion mass, the bulk away from the wall on a rectangular lattice explores a larger set of topological phases than what is accessible on a cubic lattice and the fermion spectrum on the domain wall reflects the same. For the lattice defect, despite the absence of a domain wall in the fermion mass, there can be chiral modes along the wall across which the transverse lattice spacing $a_s$ changes abruptly again reflecting the bulk C.S. levels on the two sides of the discontinuity. Interestingly, for some values of the anisotropy, the number of zero mode solutions to the equations of motion does not reflect the number of topologically protected zero modes. More specifically, for a certain range of the anisotropy, there exist zero mode solutions to the equation of motion of more than one chirality. All of these modes cannot be topologically protected as the number of topologically protected chiral modes is set by the bulk Chern-Simons levels which only reflect the net chirality on the wall.  As an example, if there are $n_+$ number of positive chirality and $n_-$ number of negative chirality zero mode solutions on the wall with $n_+>n_-$, the negative chirality zero modes pair up with $n_-$ number of positive chirality zero modes leaving $n_+-n_-$ number of topologically protected positive chirality modes on the wall. This is in contrast with the isotropic lattice, where, the equations of motion never have solutions of different chiralities realized on the same wall and all zero mode solutions to the equations of motion thus remain topologically protected.
 
There is another interesting feature associated with the lattice spacing defect considered in this paper which distinguishes it from a domain wall defect in the fermion mass. To understand this feature note that for a domain wall in the fermion mass on a uniform lattice, the C.S. current on one side of the wall where the fermion mass is negative is always zero. When there are massless modes on the domain wall (in fermion mass), the side of the wall with positive mass has nontrivial C.S. level thus supplying the entire C.S. current required to satisfy anomaly inflow constraints. In contrast with this, the lattice defect can harbor nontrivial C.S. theories on both sides of the defect, each contributing to current conservation and anomaly inflow in the presence of edge modes on the defect.

The organization of this paper is as follows. I first consider the dependence of bulk C.S. level on the lattice anisotropy. This is followed by 
an analysis of the edge modes on a domain wall for the fermion mass and a defect in lattice spacing where the lattice spacing transverse to the defect changes abruptly. I then make connections to some recent literature on Chern insulators and follow it up with a conclusion and future work. 
\section{Bulk Chern-Simons theory on a rectangular lattice}
In this section I will review the topological phases exhibited by $U(1)$ lattice gauge theory coupled to a Wilson fermion on a cubic lattice and then go on to discuss how the analysis changes on a rectangular lattice (i.e. a lattice with anisotropic lattice spacings). Along the way, I will also clarify the kind of lattice anisotropy I consider in this paper. In order to proceed consider a heavy Wilson fermion of mass $m$ and Wilson parameter $r$ coupled to a $U(1)$ lattice gauge theory in $2+1$ dimensions. I denote the lattice spacing in the direction $\mu$ as $a_{\mu}$. The Wilson-Dirac operator is then given by
\beq
D_W=\sum_{\mu=1}^3\gamma_{\mu}\partial_{\mu}+m+\frac{r}{2}\sum_{\mu=1}^3\Delta_{\mu}
\label{WD}
\eeq
where $\partial_{\mu}$ is the lattice derivative $\partial_{\mu}=\frac{\delta_{z,z+a_{\mu}}-\delta_{z,z-a_{\mu}}}{2a_{\mu}}$ and $\Delta_{\mu}$ is the lattice Laplacian $\Delta_{\mu}=\frac{\delta_{z,z+a_{\mu}}+\delta_{z,z-a_{\mu}}-2\delta_{z,z}}{a_{\mu}^2}$. Note that at this stage the lattice spacings in different directions are not specified to be equal to each other. In fact, this Wilson-Dirac operator of Eq. \ref{WD} encodes the definition of lattice anisotropy used in this paper, i.e. although the lattice spacings in the $0, 1$ and $2$ directions are free to be different from each other, the lattice Laplacian maps to the Euclidean invariant continuum Laplacian in the limit of $a_{\mu}\rightarrow 0$. Similarly the lattice derivative maps to the continuum gradient in the limit $a_{\mu}\rightarrow 0$. 
As stated earlier, for isotropic lattices, i.e. $a_0=a_1=a_2=a_{l}$, varying the dimensionless fermion mass parameter $m a_l$ with respect to the dimensionless Wilson parameter $r/a_l$ can give rise to different topological phases. A relatively simple way to see this at weak coupling is to integrate out the Wilson fermion to arrive at the low energy effective theory for the $U(1)$ gauge field. Just as in the continuum analysis with a heavy Dirac fermion and a Pauli-Villars(PV) regulator, integrating out a heavy Wilson fermion with lattice regularization in $2+1$ dimensions may result in a Chern-Simons theory at low energies. However there are some crucial difference between the lattice and PV regulator. In the continnum with PV regulator, the Chern-Simons level obtained after integrating out a heavy Dirac fermion is given by $\frac{1}{8\pi}\frac{m}{|m|}+\frac{1}{8\pi}\frac{\Lambda}{|\Lambda|}$ where $\Lambda$ is the mass of the PV field. In contrast, the Chern-Simons level obtained on a lattice after integrating out the heavy Wilson fermion is zero for $\frac{m a_l^2}{r}>6$ and $\frac{m a_l^2}{r}<0$. That the C.S. level goes to zero for certain values of the parameters $m$ and $r$ can be attributed to the presence of doublers which result from lattice regularization of the theory. For $6>\frac{m a_l^2}{r}>0$, the bulk theory exhibits three topological phases separated by two topological transitions as reviewed next.

\begin{figure}[h!]
\centering
\includegraphics[width=.7\textwidth]{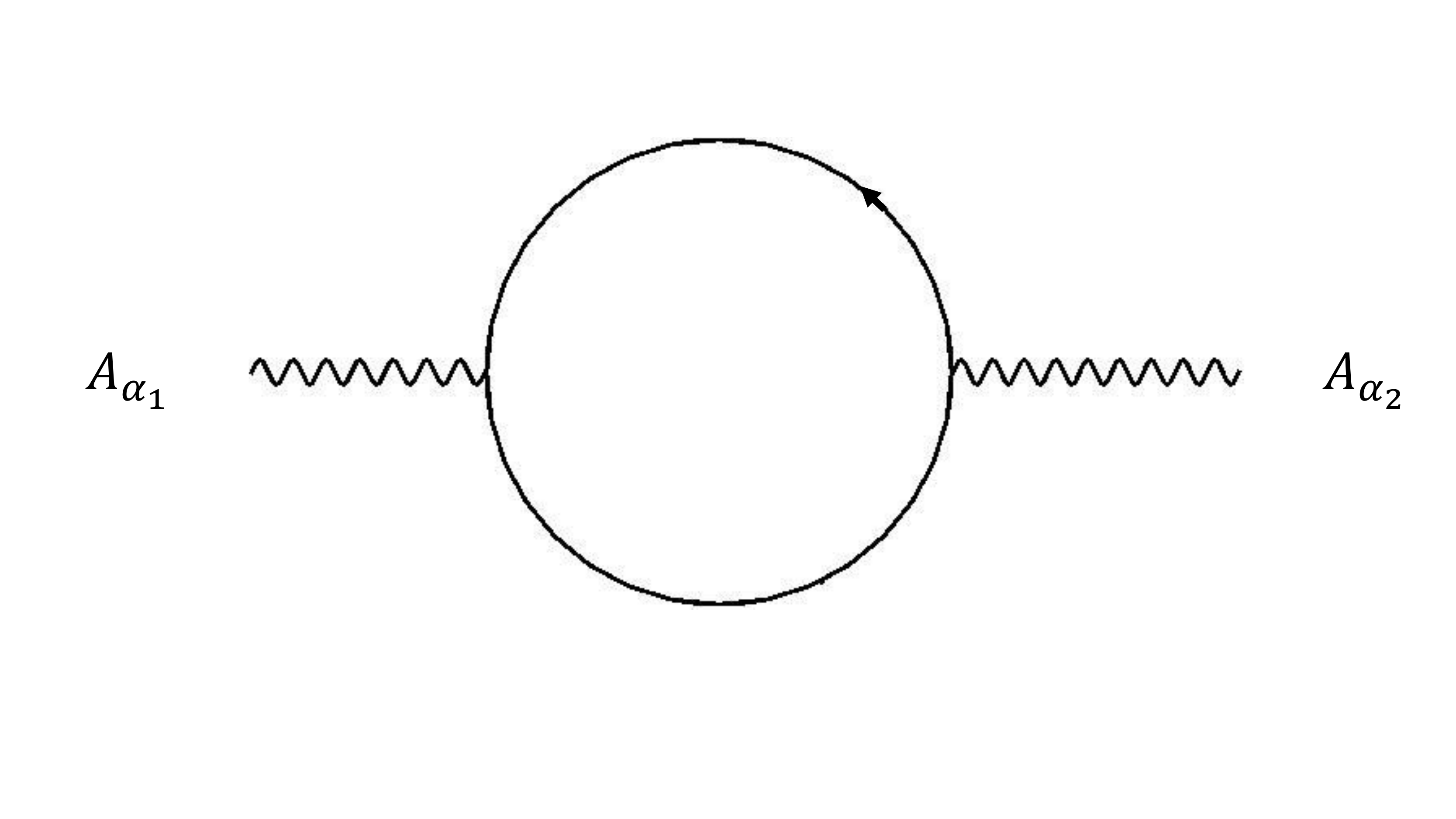}
\caption{The one-loop feynman diagram producing the Chern-Simons level.}
\label{feyn2}
\end{figure}

In order to understand how the topological transitions arise consider first the Wilson fermion propagator which is given by
\beq
S^{-1}(p)=\sum_{\mu=1}^{d} i\gamma^{\mu}\frac{\sin(p^{\mu}a^{\mu})}{a^{\mu}}+m+r\sum_{\mu=1}^{d}\frac{(\cos(p_{\mu}a^{\mu})-1)}{(a^{\mu})^2}
\label{Dfermi}
\eeq
where $d=3$ is the number of space time dimensions. In the limit of weak gauge coupling, the Wilson fermion can be integrated out which yields a Chern-Simons action for the gauge field given by $S_{\text{eff}}=-i\frac{c}{4\pi}\Gamma_{C.S.}$ with
\beq
\Gamma_{C.S.}=\epsilon_{\alpha_1\beta_1\alpha_2}\int d^3x \,\,A_{\alpha_1}\partial_{\beta_1}A_{\alpha_2}.
\eeq
The constant `$c$' is known as the Chern-Simons level which can be computed from the feynman diagram Fig. \ref{feyn2} and written as
\beq
c=-\frac{4\pi\epsilon_{\alpha_1\beta_1\alpha_2}}{2(3!)}\partial_{(q_1)_{\beta_1}}\int_{\text{BZ}}\frac{d^3p}{(2\pi)^3}\text{Tr}\left(S(p)\Lambda_{\alpha_1}(p,p-q_1)S(p-q_1)\Lambda_{\alpha_2}(p+q_2,p)\right)\big|_{q_i=0}.\nonumber\\
\eeq
Here $\Lambda$ is the fermion-Gauge field vertex satisfying
\beq
\Lambda_{\mu}(p,p)=-i\partial_{p_{\mu}}S^{-1}(p).
\eeq
The coefficient `$c$' can be re-expressed in terms of the fermion propagator as 
\beq
c=\frac{\epsilon_{\mu_1\mu_2\mu_3}}{2(3!)}\int \frac{d^3p}{2\pi^2}\text{Tr}\left(\left[S(p)\partial_{p_{\mu_1}}S^{-1}(p)\right]\left[S(p)\partial_{p_{\mu_2}}S^{-1}(p)\right]\left[S(p)\partial_{p_{\mu_3}}S^{-1}(p)\right]\right).\nonumber\\
\label{cn2}
\eeq
One can then substitute the Wilson fermion propagator in Eq. \ref{cn2} to evaluate `$c$' by computing the momentum space integral near the Brillouin zone (BZ) corners, the coordinates of which are denoted as $\xi_{\alpha}$. Here $\alpha=1,..,\binom{d}{k}$ where $k$ stands for the number of components of the momenta equal to $\pi$ while the rest of the components are zero. The C.S. can then be written as
\beq
c=\sum_{k,\alpha}\int_{d\Omega}(-1)^k\frac{d^3p}{2\pi^2}\frac{(m-2\frac{r}{a_l^2}k)}{\left(p^2+(m-2\frac{r}{a_l^2}k)^2\right)^2}.
\eeq
Here I have assumed $a_0=a_1=a_2=a_l$.
Performing this integral results in the following formula for the CS coefficient
\beq
c&=&\sum_{k=0}^d (-1)^k \binom{d}{k} \frac{(m-2\frac{r}{a_l^2}k)}{|m-2\frac{r}{a_l^2}k|}\nonumber\\
&=&\sum_{k=0}^d (-1)^k \binom{d}{k} \frac{(M-2Rk)}{|M-2Rk|}\nonumber\\
\eeq
where I have used $M=m a_l$ and $R=r/a_l$ as before.
It is now straightforward to check that for $0<M/R<2$, $c=-1$, for $2<M/R<4$, 
$c=2$ and for $4<M/R<6$, 
$c=-1$. Similarly, for $M/R>6$ and for $M/R<0$, $c=0$. Each of the different C.S. levels obtained here corresponds to a distinct topological phase. 
I will now consider a rectangular lattice, in particular with $a_0=a_1=a$ and $a_2=a_s$ such that $a\neq a_s$. In this case the Chern-Simons coefficient can be obtained by substituting fermion propagators from Eq. \ref{Dfermi} in Eq. \ref{cn2} and repeating the procedures outlined above. The C.S. level is then given by 
\beq
c&=&\frac{1}{2}\left[(-1)^0 \frac{m}{|m|}+(-1)^1\left(\frac{m-\frac{2r}{a^2}}{|m-\frac{2r}{a^2}|}2+\frac{m-\frac{2r}{a_s^2}}{|m-\frac{2r}{a_s^2}|}\right)\right.\nonumber\\
&+&\left.(-1)^2\left(\frac{m-\frac{4r}{a^2}}{|m-\frac{4r}{a^2}|}+2\frac{m-\frac{2r}{a^2}-2\frac{r}{a_s^2}}{|m-\frac{2r}{a^2}-2\frac{r}{a_s^2}|}\right)\right.\nonumber\\
&+&\left.(-1)^3\left(\frac{m-4\frac{r}{a^2}-2\frac{r}{a_s^2}}{|m-4\frac{r}{a^2}-2\frac{r}{a_s^2}|}\right)\right].
\eeq
It is now clear that the C.S. level depends on the lattice spacings $a_s$ and $a$ and will take various different values in integers as the two lattice spacings are varied. Each of these different C.S. levels will correspond to a distinct topological phase. In other words, a rectangular lattice is able to access a larger set of topological phases than a cubic lattice. In the next section
I analyze the Chern-Simons level as a function of the lattice anisotropy and some of the other parameters in the theory. This analysis will pave the way for understanding the edge modes on the two kinds of defects that I will consider subsequently: a domain wall in fermion mass and an abrupt change in lattice spacing.
\subsection{Variation in Chern-Simons level for various values of lattice anisotropy as a function of fermion mass}
\label{CS1}
\begin{figure}[h!]
\centering
\includegraphics[width=.7\textwidth]{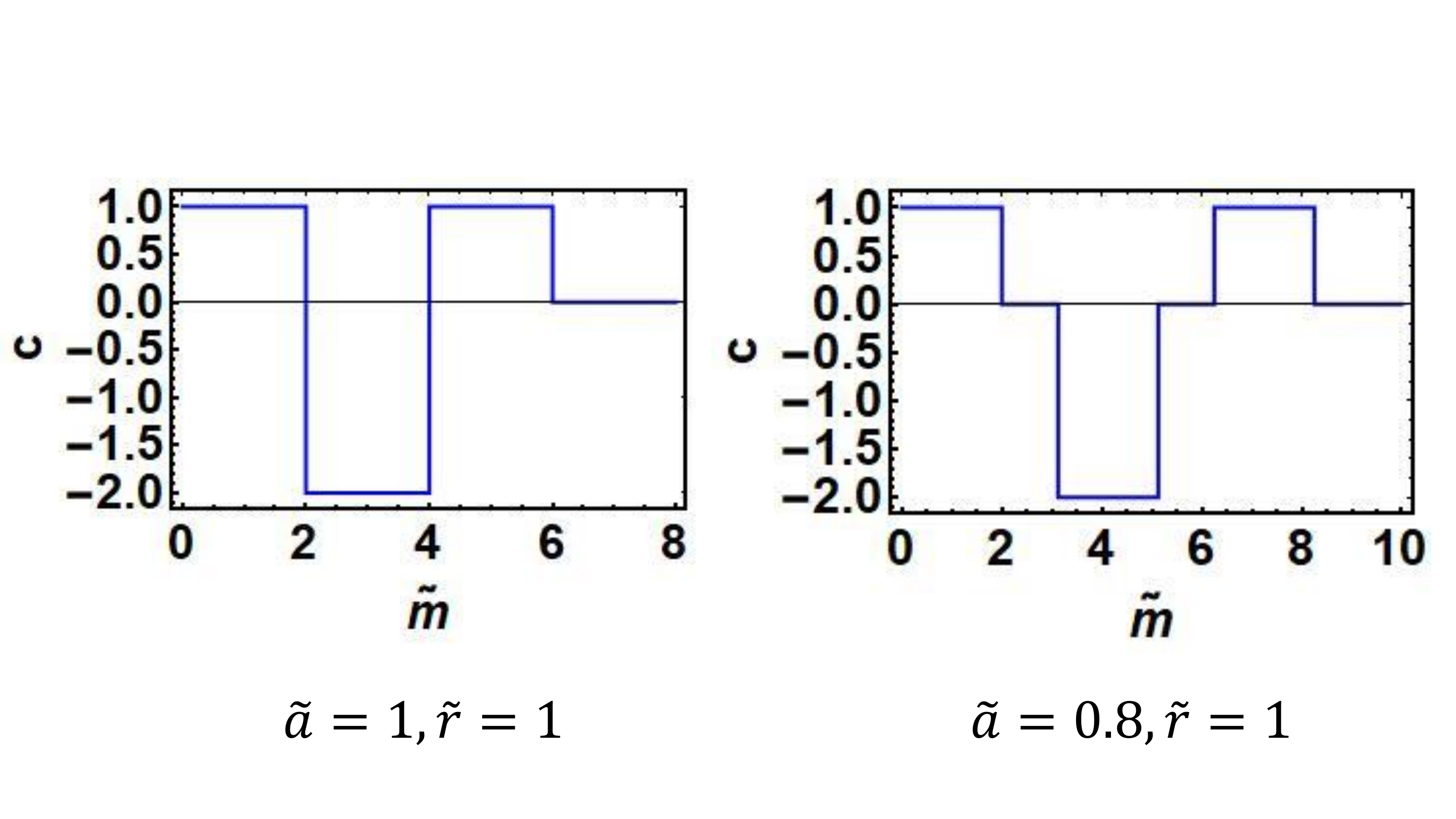}
\caption{Chern-Simons level as a function of $\tilde{m}$ with $\tilde{r}=1$ for various values of lattice anisotropy.}
\label{atv1}
\end{figure}

\begin{figure}[h!]
\centering
\includegraphics[width=.7\textwidth]{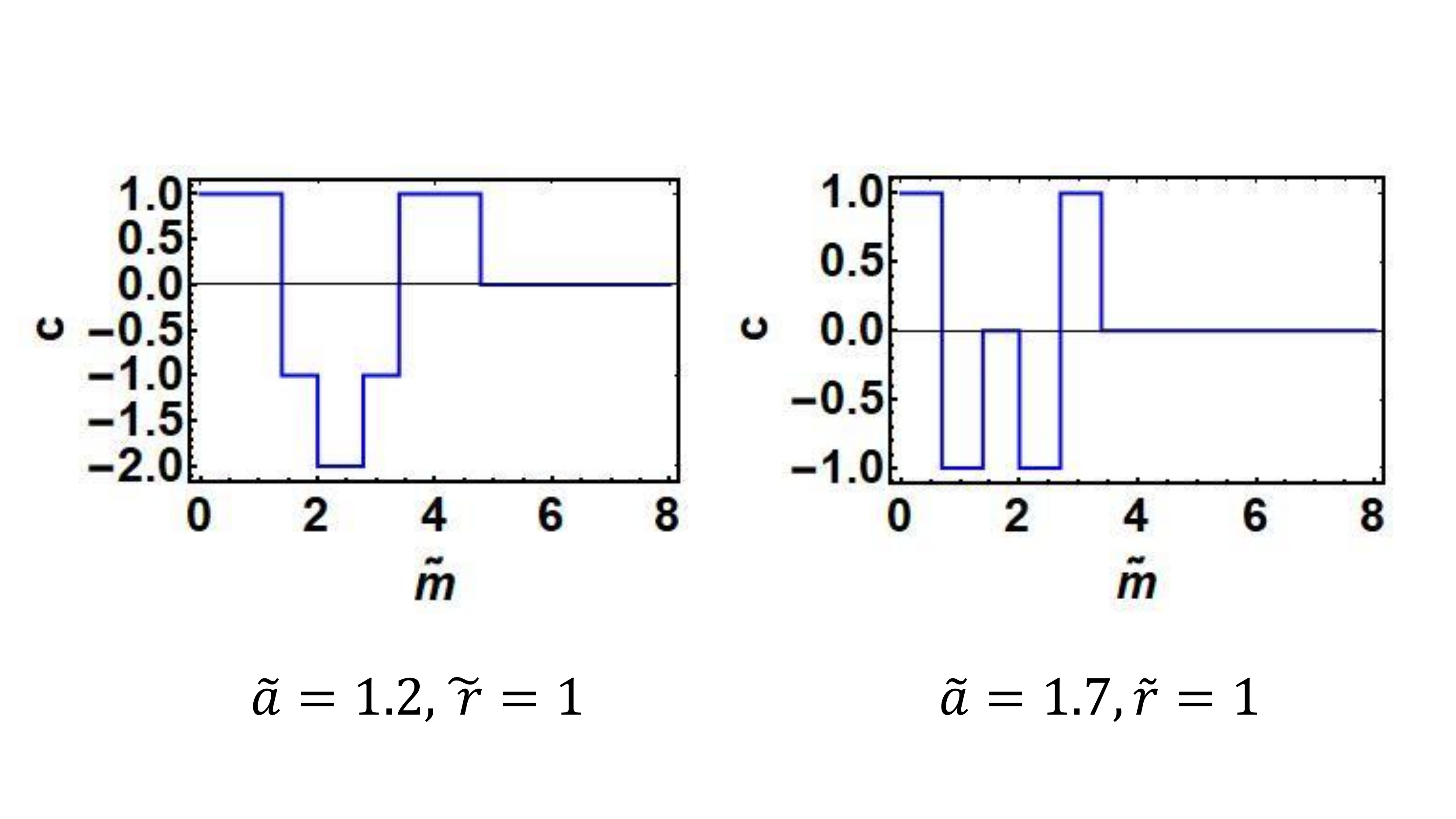}
\caption{Chern-Simons level as a function of $\tilde{m}$ with $\tilde{r}=1$ for various values of lattice anisotropy.}
\label{atv2}
\end{figure}

I will first consider the behavior of the C.S. level with respect to variations in the fermion mass $m$ for a fixed lattice anisotropy $a_s/a$. To do this I define the dimensionless parameters $m a_s=\tilde{m}$, $\frac{r}{a_s}=\tilde{r}$ and use the lattice anisotropy parameter $\tilde{a}=\frac{a}{a_s}$ to rewrite the C.S. level as
\beq
c&=&\frac{1}{2}\left[(-1)^0\frac{\tilde{m}}{|\tilde{m}|}+(-1)^1\left(2\frac{(\tilde{m}-2\frac{\tilde{r}}{\tilde{a}^2})}{\big|\tilde{m}-2\frac{\tilde{r}}{\tilde{a}^2}\big|}+\frac{(\tilde{m}-2\tilde{r})}{\big|\tilde{m}-2\tilde{r}\big|}\right)\right.\nonumber\\
&&\left.+(-1)^2\left(\frac{(\tilde{m}-4\frac{\tilde{r}}{\tilde{a}^2})}{\big|\tilde{m}-4\frac{\tilde{r}}{\tilde{a}^2}\big|}+2\frac{(\tilde{m}-2\frac{\tilde{r}}{\tilde{a}^2}-2\tilde{r})}{\big|\tilde{m}-2\frac{\tilde{r}}{\tilde{a}^2}-2\tilde{r}\big|}\right)+(-1)^3\frac{(\tilde{m}-4\frac{\tilde{r}}{\tilde{a}^2}-2\tilde{r})}{\big|\tilde{m}-4\frac{\tilde{r}}{\tilde{a}^2}-2\tilde{r}\big|}\right].\nonumber\\
\eeq
I now set $\tilde{r}=1$ and plot the C.S. level `$c$' as a function of $\tilde{m}$ for a few different values of $\tilde{a}$ in Fig. \ref{atv1} and \ref{atv2}. Note that $c=0$ for $\tilde{m}<0$ for all values of the lattice anisotropy and is not plotted hence. As shown in Fig. \ref{atv1} for $\tilde{a}=1$, I recover the results for an isotropic lattice i.e. `$c$' goes between $ 1, -2, 1, 0$ for $0<\tilde{m}<2$, $2<\tilde{m}<4$, $4<\tilde{m}<6$ and $6<\tilde{m}$ respectively. For $\tilde{a}<1$, the regions of parameter space in $\tilde{m}$ where the C.S. level changes between $1,-2, 1$ get separated from each other by regions where `$c$' is just zero. In other words the C.S. level takes values $0$ for $\tilde{m}<0$, $1$ in the region $2>\tilde{m}>0$, $0$ in $\frac{2}{\tilde{a}^2}>\tilde{m}>2$, $-2$ in $2+\frac{2}{\tilde{a}^2}>\tilde{m}>\frac{2}{\tilde{a}^2}$, $0$ in $\frac{4}{\tilde{a}^2}>\tilde{m}>2+\frac{2}{\tilde{a}^2}$, $1$ in $2+\frac{4}{\tilde{a}^2}>\tilde{m}>\frac{4}{\tilde{a}^2}$ and $0$ in $\tilde{m}>2+\frac{4}{\tilde{a}^2}$. To illustrate these behaviors I plot `$c$' as a function of $\tilde{m}$ in Fig. \ref{atv1} for $\tilde{a}=1$ and $\tilde{a}=0.8$.\\
For $\tilde{a}>1$, the behavior of `$c$' as a function of $\tilde{m}$ for $\tilde{r}=1$ exhibits two different patterns for $\sqrt{2}>\tilde{a}>1$ and $\tilde{a}>\sqrt{2}$ as shown in Fig. \ref{atv2}. For $\sqrt{2}>\tilde{a}>1$, `$c$' toggles between $0, 1, -1, -2, -1, 1, 0$ for $\tilde{m}<0$, $0<\tilde{m}<\frac{2}{\tilde{a}^2}$, $\frac{2}{\tilde{a}^2}<\tilde{m}<2$, $2<\tilde{m}<\frac{4}{\tilde{a}^2}$, $\frac{4}{\tilde{a}^2}<\tilde{m}<2+\frac{2}{\tilde{a}^2}$
and $2+\frac{2}{\tilde{a}^2}<\tilde{m}<2+\frac{4}{\tilde{a}^2}$ respectively. For $\tilde{a}>\sqrt{2}$, on the other hand, `$c$' jumps between $0, 1, -1, 0, -1, 1$ for $\tilde{m}<0$, $0<\tilde{m}<\frac{2}{\tilde{a}^2}$, $\frac{2}{\tilde{a}^2}<\tilde{m}<\frac{4}{\tilde{a}^2}$,
$\frac{4}{\tilde{a}^2}<\tilde{m}<2$, $2<\tilde{m}<2+\frac{2}{\tilde{a}^2}$, $2+\frac{2}{\tilde{a}^2}<\tilde{m}<2+\frac{4}{\tilde{a}^2}$ and $\tilde{m}>2+\frac{4}{\tilde{a}^2}$. The C.S. levels computed in this subsection will be useful in the next section when I investigate the edge modes localized on a domain wall in the fermion mass for a uniform rectangular lattice.  
\subsection{Variation in Chern-Simons level as a function of lattice anisotropy for various values of fermion mass}
\label{sub2}
\begin{figure}[h!]
\centering
\includegraphics[width=\textwidth]{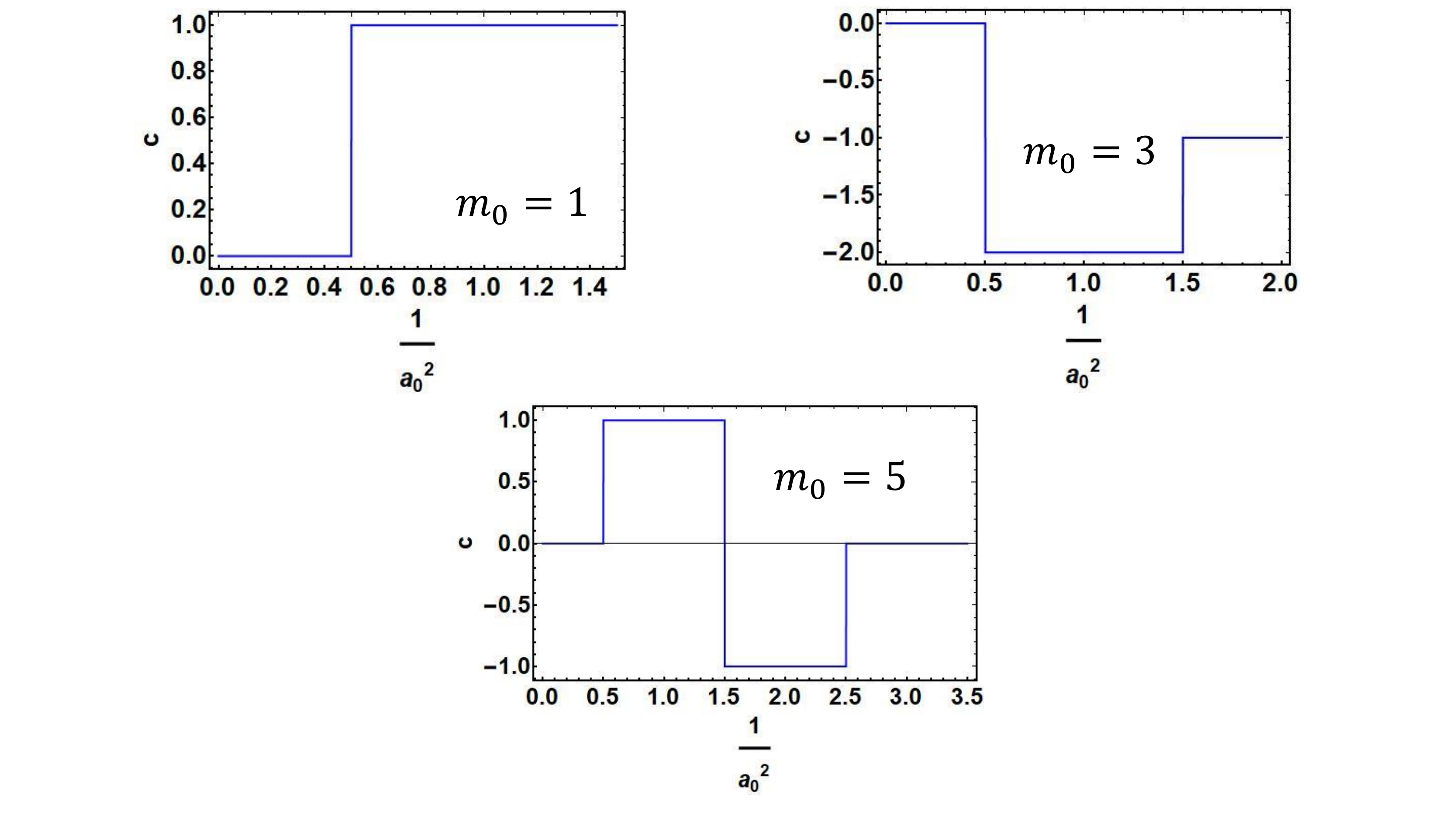}
\caption{Chern-Simons level as a function of lattice anisotropy for various values of $m_0$.}
\label{fig1}
\end{figure}

Next, I consider the behavior of the C.S. level with variation in lattice anisotropy for various values of $m$ (appropriately normalized) while keeping $r$ fixed (appropriately normalized). For this analysis I will define the following dimensionless variables $m_0=m a$, $r_0=\frac{r}{a}$ and $\frac{a_s}{a}=a_0$. In terms of these variables the C.S. level `$c$' can be written as
\beq
c&=&\frac{1}{2}\left[(-1)^0\frac{m_0}{|m_0|}+(-1)\left(\frac{m_0-2\frac{r_0}{a_0^2}}{|m_0-2\frac{r_0}{a_0^2}|}+2\frac{m_0-2r_0}{|m_0-2r_0|}\right)\right.\nonumber\\
&+&\left.(-1)^2 \left(2\frac{m_0-r_0(2+\frac{2}{a_0^2})}{|m_0-r_0(2+\frac{2}{a_0^2})|}+\frac{m_0-4r_0}{|m_0-4r_0|}\right)+(-1)^3\left(\frac{m_0-r_0(\frac{2}{a_0^2}+4)}{|m_0-r_0(\frac{2}{a_0^2}+4)|}\right)\right].\nonumber\\
\eeq
I fix $r_0=1$ and plot `$c$' as a function of $\frac{1}{a_0^2}$ for three different values of $m_0=1, 3, 5$ in Fig. \ref{fig1}. 

As shown in Fig. \ref{fig1} for $a_0=1$, I recover the C.S. levels for an isotropic lattice. More generally, for $m_0=1$ the C.S. level is $0$ for $\frac{1}{a_0^2}<\frac{m_0}{2}$ and is $-1$ for $\frac{1}{a_0^2}>\frac{m_0}{2}$. For $m_0=3$, the C.S. level is $0, 2$ and $1$ in the regions $\frac{1}{a_0^2}<\frac{m_0}{2}-1$, $\frac{m_0}{2}-1<\frac{1}{a_0^2}<\frac{m_0}{2}$ and $\frac{1}{a_0^2}>\frac{m_0}{2}$ respectively. Similarly, for $m_0=5$, the C.S. level alternates between $0, -1, 1, 0$ for $\frac{1}{a_0^2}<\frac{m_0}{2}-2$, $\frac{m_0}{2}-2<\frac{1}{a_0^2}<\frac{m_0}{2}-1$, $\frac{m_0}{2}-1<\frac{1}{a_0^2}<\frac{m_0}{2}$ and $\frac{1}{a_0^2}>\frac{m_0}{2}$ respectively. Again, the variation of the C.S. level as a function of the lattice anisotropy $\frac{1}{a_0^2}$ establishes that the anisotropy parameter $a_0$, when dialed can drive the lattice theory to different topological phases. The C.S. levels computed here as a function of $\frac{1}{a_0^2}$ will help in understanding the existence of topologically protected edge modes from the perspective of anomaly inflow when I consider an abrupt change in the lattice spacing in section \ref{ablat}. 

With this I now proceed to analyze the edge modes on a domain wall in fermion mass. This will be followed by the analysis of edge modes on a defect across which lattice spacing changes abruptly while all other parameters of the theory are kept fixed.
\section{Domain wall in fermion mass on a rectangular lattice}
\begin{figure}[h!]
\centering
\includegraphics[width=.7\textwidth]{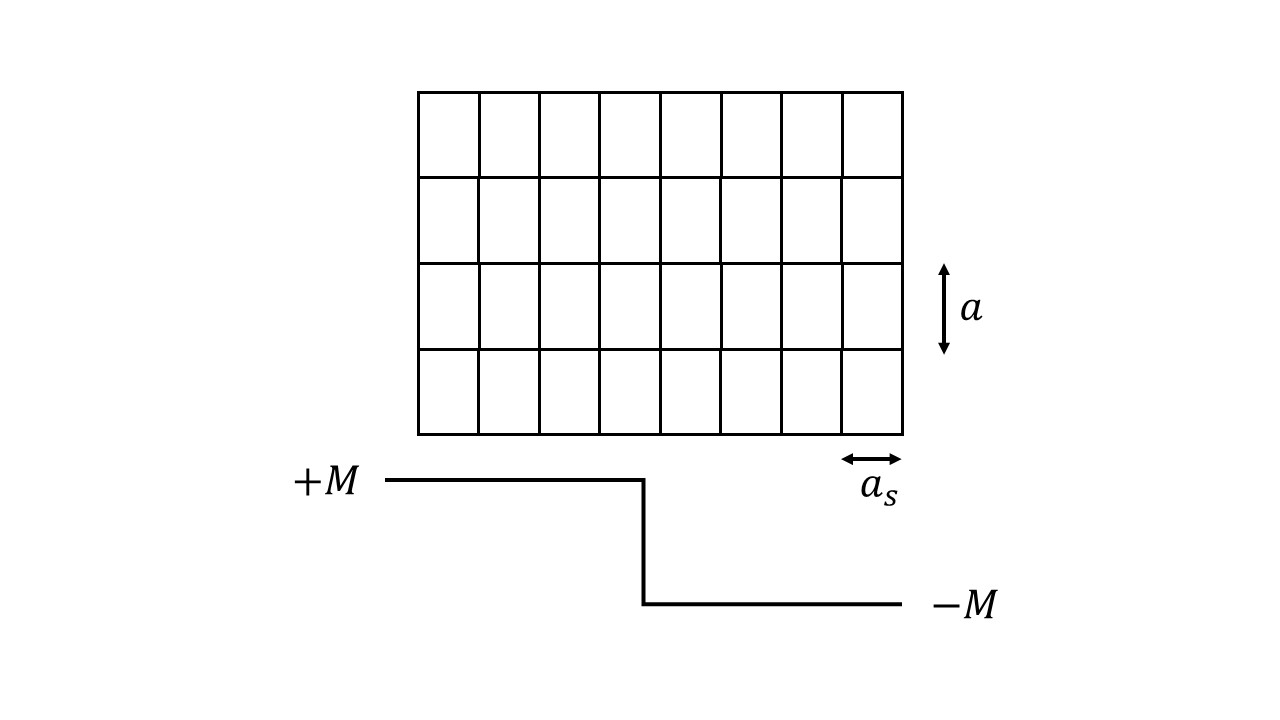}
\caption{Anisotropic lattice with domain wall in fermion mass.}
\label{fig_lat1}
\end{figure}

In this section I investigate the existence of chiral edge modes in the presence of a domain wall in the fermion mass parameter on an anisotropic, but uniform lattice as shown in Fig. \ref{fig_lat1}. To do so I consider the Dirac equation with a spatially dependent mass of the form $m(x_2)=\mu\epsilon(x_2)$ with $\mu>0$ such that there is a domain wall at $x_2=0$. With lattice spacings $a$ in the $x_0$ and $x_1$ directions and $a_s$ in the $x_2$ direction
the Dirac equation takes the form
\beq
\pm\frac{\phi_{\pm}(x_2+a_s)-\phi_{\pm}(x_2-a_s)}{2a_s}&+&m(x_2)\phi_{\pm}(x)+\frac{r}{2}\sum_{i=0,1}\left(\frac{\phi_{\pm}(x_i+a)+\phi_{\pm}(x_i-a)-2\phi_{\pm}(x_i)}{a^2}\right)\nonumber\\\nonumber\\
&&\,\,\,\,\,\,\,\,\,\,\,\,\,\,+\frac{r}{2}\left(\frac{\phi_{\pm}(x_2+a_s)+\phi_{\pm}(x_2-a_s)-2\phi_{\pm}(x_2)}{a_s^2}\right)=0\nonumber\\
\eeq
for the two chiralities `$\pm$' where on the wall I impose,
\beq 
\gamma_i\sum_{i=0,1}\partial_i\phi=0\,\,\,\,\, \text{with}\,\,\,\, \phi=\begin{pmatrix}\phi_+\\
\phi_-
\end{pmatrix}.
\eeq
Defining $\phi_{\pm}(x_2,p)=\int \phi_{\pm}(x_2,x_i)e^{i p_i x_i}d^2x$ the Dirac equation can be written as 
\beq
\pm\frac{\phi_{\pm}(x_2+a_s,p)-\phi_{\pm}(x_2-a_s,p)}{2a_s}&+&m\epsilon(s)\phi_{\pm}(x_2,p)+\frac{r}{2}\sum_{i=0,1}\frac{\left(\cos(p_ia)-1\right)}{a^2}\phi_{\pm}(x_2,p)\nonumber\\
&+&
\frac{r}{2}\left(\frac{\phi_{\pm}(x_2+a_s,p)+\phi_{\pm}(x_2-a_s,p)-2\phi_{\pm}(x_2,p)}{a_s^2}\right)=0.\nonumber\\
\eeq
I now take $r=a_s$. In terms of the dimensionless variable $m a_s=\tilde{m}$ the equation of motion simplifies to
\beq
\phi_{\pm}(x_2\pm a_s)=-\tilde{m}_{\text{eff}}\phi_{\pm}(x_2)
\eeq
such that $\tilde{m}_{\text{eff}}=\tilde{m}(s)-1-F(p)\frac{a_s^2}{a^2}$ with $F(p)\equiv\sum_{i=0,1}(1-\cos(p_i a))$.
There are no normalizable solutions for the `$-$' chirality mode. However, `$+$' chirality mode can have normalizable solutions given by
\beq
\phi_+(x_2)=(-\tilde{m}_{\text{eff}})^{x_2/a_s}.
\label{sol_DWmass}
\eeq
For $x_2<0$, this solution is normalizable for all values of $\tilde{\mu}$. For $x_2>0$ the solution is normalizable 
only when 
\beq
2>\left(\tilde{\mu}-\left(\frac{1}{\tilde{a}}\right)^2F(p)\right)>0
\eeq 
where I have used $\tilde{a}=\frac{a}{a_s}$.
Zero mode solutions centered around $\{p_0 a=0, p_1 a=0\}$ and $\{p_0 a=\pi, p_1 a=\pi\}$ are of positive chirality and those centered around $\{p_0 a=0, p_1 a=\pi\}$ and $\{p_0 a=\pi, p_1 a=0\}$ are of negative chirality. The positive chirality mode centered at $\{p_0 a=0, p_1 a=0\}$ is normalizable for
\beq
0<\tilde{\mu}<2.
\eeq
The negative chirality modes at $\{p_0 a=\pi, p_1 a=0\}$ and $\{p_0 a=0, p_1 a=\pi\} $ are normalizable for 
\beq
\frac{2}{\tilde{a}^2}<\tilde{\mu}<2+\frac{2}{\tilde{a}^2}.
\eeq
Finally, the positive chirality mode centered at $\{p_0 a=\pi, p_1 a=\pi\}$ is normalizable for
\beq
\frac{4}{\tilde{a}^2}<\tilde{\mu}<2+\frac{4}{\tilde{a}^2}.
\eeq
With the conditions of normalizability at hand, I can now discuss the pattern of variation of the number and chirality of the zero modes as a function of $\tilde{\mu}$ for fixed $\tilde{a}$. For $\tilde{a}=1$, I recover the results for a cubic lattice where one finds one positive chirality mode for $0<\tilde{\mu}<2$, two negative chirality modes for $2<\tilde{\mu}<4$ and one positive chirality mode $4<\tilde{\mu}<6$. 
For $\tilde{a}\neq 1$, three different patterns emerge for $\tilde{a}<1$, $\sqrt{2}>\tilde{a}>1$ and $\tilde{a}>\sqrt{2}$ respectively. I discuss these patterns in the rest of this section and summarize them in tables \ref{t1}, \ref{t2} and \ref{t3}. It is instructive to note that the patterns are consistent with the C.S. levels obtained in subsection \ref{CS1} as described in Fig. \ref{atv1} and \ref{atv2}.
\\
\subsection{Varying fermion mass when the anisotropy parameter is less than one}
For $\tilde{a}<1$, the edge mode chirality on the wall alternates between the values $0, 1, 0, -2$, $0, 1, 0$ as a function of $\tilde{\mu}$. The range of values that the net chirality on the wall can take is the same as in the case with $\tilde{a}=1$, except that the three regions in the parameter space of $\tilde{\mu}$ where the net chirality takes values $1$, $-2$ and $1$ are separated from each other by regions in the parameter space with no normalizable edge mode solutions. This of course is consistent with the change in C.S. levels seen in Fig. \ref{atv1} for $\tilde{a}=0.8$. More specifically, in the region $2>\tilde{\mu}>0$, the only mode with normalizable solution is $\{p_0 a=0, p_1 a=0\}$. Similarly, in the region $2+\frac{2}{\tilde{a}^2}>\tilde{\mu}>\frac{2}{\tilde{a}^2}$ the modes with normalizable solutions are $\{p_0 a=0, p_1 a=\pi\}$ and $\{p_0 a=\pi, p_1 a=0\}$. In $2+\frac{4}{\tilde{a}^2}>\tilde{\mu}>\frac{4}{\tilde{a}^2}$ it is only the mode $\{p_0 a=\pi, p_1 a=\pi\}$ which has normalizable solution. In the regions $0>\tilde{\mu}$, $\frac{2}{\tilde{a}^2}>\tilde{\mu}>2$, $\frac{4}{\tilde{a}^2}>\tilde{\mu}>2+\frac{2}{\tilde{a}^2}$ and $\tilde{\mu}>2+\frac{2}{\tilde{a}^2}$ there are no normalizable zero mode solutions at all. I summarize the results in table \ref{t1}. 
\begin{table}[t]
 \begin{tabular}{||c| c| c| c||} 
 \hline
 Range of $\tilde{\mu}$ & Positive chirality & Negative chirality & Net chirality \\ [0.5ex] 
 \hline\hline
 $2>\tilde{\mu}>0$ & 1 & 0 & +1\\ 
 \hline
 $\frac{2}{\tilde{a}^2}>\tilde{\mu}>2$ & 1 & 1 & 0\\
 \hline
 $2+\frac{2}{\tilde{a}^2}>\tilde{\mu}>\frac{2}{\tilde{a}^2}$ & 0 & 2 & -2 \\
 \hline
 $\frac{4}{\tilde{a}^2}>\tilde{\mu}>2+\frac{2}{\tilde{a}^2}$ & 1 & 1 & 0\\
 \hline
 $2+\frac{4}{\tilde{a}^2}>\tilde{\mu}>\frac{4}{\tilde{a}^2}$ & 1 & 0 & 1\\ [1ex] 
 \hline
\end{tabular}
\caption{Summary of the zero mode solutions on the domain wall as a function of $\tilde{\mu}$ for $\tilde{a}<1$.}
\label{t1}
\end{table}\\
\subsection{Varying fermion mass when the anisotropy parameter is greater than one}
For $\tilde{a}>1$ something even more interesting takes place. For certain values of the lattice anisotropy parameter, there can exist normalizable solutions of different chiralities on the domain wall. Of course in that case, all of the normalizable solutions are not topologically protected and some linear combinations of these modes will acquire a nonzero mass when interactions are taken into account. As a result the number of topologically protected massless modes on the wall equals the net chirality present on the wall which does not coincide with the number of zero mode solutions to the Dirac equation. The parameter space in $\tilde{a}>1$ exhibits two different patterns for the variation of the edge modes and their chirality for $\tilde{a}>\sqrt{2}$ and $\tilde{a}<\sqrt{2}$ as mentioned earlier which I now discuss.\\
\subsubsection{Variation with fermion mass for $\sqrt{2}>\tilde{a}>1$}
In this case, for $\frac{2}{\tilde{a}^2}>\tilde{\mu}>0$, it is only
$\{p_0 a=0,p_1 a=0\}$ which is normalizable. For $2>\tilde{\mu}>\frac{2}{\tilde{a}^2}$, the normalizable modes are $\{p_0 a=0,p_1 a=0\}$, $\{p_0 a=0,p_1 a=\pi\}$ and $\{p_0 a=\pi,p_1 a=0\}$. Similarly, in the region $\frac{4}{\tilde{a}^2}>\tilde{\mu}>2$ the only normalizable modes are $\{p_0 a=0,p_1 a=\pi\}$ and $\{p_0 a=\pi,p_1 a=0\}$. For $2+\frac{2}{\tilde{a}^2}>\tilde{\mu}>\frac{4}{\tilde{a}^2}$ normalizable solutions exist for $\{p_0 a=0,p_1 a=\pi\}$, $\{p_0 a=\pi,p_1 a=0\}$ and $\{p_0 a=\pi,p_1 a=\pi\}$. Finally, for $2+\frac{4}{\tilde{a}^2}>\tilde{\mu}>2+\frac{2}{\tilde{a}^2}$ it is only $\{p_0 a=\pi,p_1 a=\pi\}$ which has normalizable solution. No normalizable solutions exist for any of the modes when $\tilde{\mu}<0$ or $\tilde{\mu}>2+\frac{4}{\tilde{a}^2}$. The normalizable zero mode solutions and their net chirality are summarized in table \ref{t2}.\\
\subsubsection{Variation with fermion mass for $\tilde{a}>\sqrt{2}$}
For $\tilde{a}>\sqrt{2}$ zero modes are realized on the domain wall in the following pattern. For $\tilde{\mu}<0$ and $\tilde{\mu}>2+\frac{4}{\tilde{a}^2}$ there are no normalizable zero mode solutions. For $\frac{2}{\tilde{a}^2}>\tilde{\mu}>0$, only $\{p_0 a=0, p_1 a=0\}$ has normalizable solution. For $\frac{4}{\tilde{a}^2}>\tilde{\mu}>\frac{2}{\tilde{a}^2}$ normalizable solutions exist for the modes $\{p_0 a=0, p_1a=0\}$, $\{p_0 a=0, p_1 a=\pi\}$ and $\{p_0 a=\pi,p_1 a=0\}$. All four modes $\{p_0 a=0, p_1 a=0\}$, $\{p_0 a=0, p_1 a=\pi\}$, $\{p_0 a=\pi, p_1 a=0\}$, $\{p_0 a=\pi, p_1 a=\pi\}$ have normalizable solutions in $2>\tilde{\mu}>\frac{4}{\tilde{a}^2}$ resulting in no net chirality on the wall. Between $2+\frac{2}{\tilde{a}^2}>\tilde{\mu}>2$ the normalizable zero modes are $\{p_0 a=0, p_1 a=\pi\}$, $\{p_0 a=\pi, p_1 a=0\}$ and $\{p_0 a=\pi, p_1 a=\pi\}$. Finally, in $2+\frac{4}{\tilde{a}^2}>\tilde{\mu}>2+\frac{2}{\tilde{a}^2}$ normalizable solution exists only for the mode $\{p_0 a=\pi, p_1 a=\pi\}$. The normalizable zero mode solutions and their net chirality are summarized in table \ref{t3}.

To see that the zero mode solutions obtained here are consistent with the anomaly inflow constraints, note that the net chirality of zero modes listed in tables \ref{t1}, \ref{t2} and \ref{t3} coincide with the C.S. levels in Fig. \ref{atv1} and \ref{atv2}. Also, note that on one side of the domain wall where the fermion mass is positive, the C.S. level is given by Fig. \ref{atv1} and \ref{atv2}. The other side of the wall $x_2<0$ with negative fermion mass has C.S. level of zero. Thus there is no current flowing to the wall from $x_2<0$ and the current flowing on the $x_2>0$ bulk matches the net chirality on the wall exactly.

\begin{table}[t]
 \begin{tabular}{||c| c| c| c||} 
 \hline
 Range of $\tilde{\mu}$ & Positive chirality & Negative chirality & Net chirality \\ [0.5ex] 
 \hline\hline
 $2\left(\frac{1}{\tilde{a}}\right)^2>\tilde{\mu}>0$ & 1 & 0 & +1\\ 
 \hline
 $2>\tilde{\mu}>2\left(\frac{1}{\tilde{a}}\right)^2$ & 1 & 2 & -1\\
 \hline
 $4\left(\frac{1}{\tilde{a}}\right)^2>\tilde{\mu}>2$ & 0 & 2 & -2 \\
 \hline
 $2+2\left(\frac{1}{\tilde{a}}\right)^2>\tilde{\mu}>4\left(\frac{1}{\tilde{a}}\right)^2$ & 1 & 2 & -1\\
 \hline
 $2+4\left(\frac{1}{\tilde{a}}\right)^2>\tilde{\mu}>2+2\left(\frac{1}{\tilde{a}}\right)^2$ & 1 & 0 & 1\\ [1ex] 
 \hline
\end{tabular}
\caption{Summary of the zero mode solutions on the domain wall as a function of $\tilde{\mu}$ for $\sqrt{2}>\tilde{a}>1$.}
\label{t2}
\end{table}

\begin{table}[t]
 \begin{tabular}{||c| c| c| c||} 
 \hline
 Range of $\tilde{\mu}$ & Positive chirality & Negative chirality & net chirality \\ [0.5ex] 
 \hline\hline
 $2\left(\frac{1}{\tilde{a}}\right)^2>\tilde{\mu}>0$ & 1 & 0 & +1\\ 
 \hline
 $4\left(\frac{1}{\tilde{a}}\right)^2>\tilde{\mu}>2\left(\frac{1}{\tilde{a}}\right)^2$ & 1 & 2 & -1\\
 \hline
 $2>\tilde{\mu}>4\left(\frac{1}{\tilde{a}}\right)^2$ & 2 & 2 & 0 \\
 \hline
 $2+2\left(\frac{1}{\tilde{a}}\right)^2>\tilde{\mu}>2$ & 1 & 2 & -1\\
 \hline
 $2+4\left(\frac{1}{\tilde{a}}\right)^2>\tilde{\mu}>2+2\left(\frac{1}{\tilde{a}}\right)^2$ & 1 & 0 & 1\\ [1ex] 
 \hline
\end{tabular}
\caption{Summary of the zero mode solutions on the domain wall as a function of $\tilde{\mu}$ for $\tilde{a}>\sqrt{2}$.}
\label{t3}
\end{table}

\section{Abrupt change in lattice spacing}
\label{ablat}
\begin{figure}[h!]
\centering
\includegraphics[width=.7\textwidth]{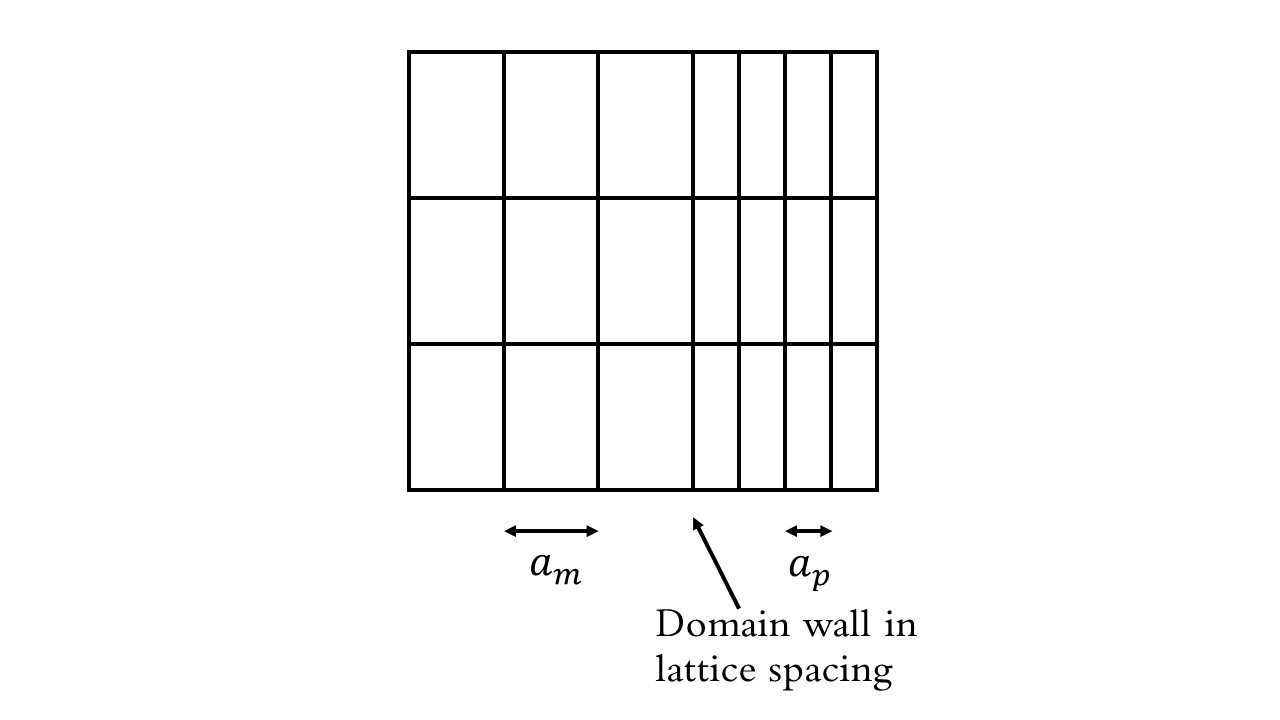}
\caption{Lattice with an abrupt change in lattice spacing.}
\label{fig_lat2}
\end{figure}

In this section I will analyze a Wilson fermion with no spatial variation in its mass formulated on a lattice with a domain wall defect in the lattice spacing as shown in Fig. \ref{fig_lat2}. The lattice is such that, the spacing in $x_0$ and $x_1$ directions, i.e. $a$ is fixed. However lattice spacing in $x_2$ direction $a_2$ changes discontinuously across $x_2=0$ as shown in Fig. \ref{fig_lat2} according to
\beq
a_2(x_2)=a_p\theta(x_2)+a_m\theta(-x_2)
\eeq
where $\theta(x)$ is the unit step function.
All the other parameters in the theory i.e. the Wilson parameter, fermion mass and the lattice spacing in the $x_0$ and $x_1$ directions are kept constant. Quite remarkably, I find that there exist topologically protected chiral zero mode solutions for certain values of the parameters $a_p$ and $a_m$ despite there being no domain wall in the fermion mass. Another related feature of such a defect is that there exist $a_p$ and $a_m$ for which the C.S. levels on both sides of the defect are nonzero as opposed a domain wall defect in the fermion mass where the C.S. level on one side of the wall is always zero. 

 In order to find zero mode solutions on the defect one has to write down Dirac equation for the lattice in Fig. \ref{fig_lat2}. For this purpose it is convenient to use the following dimensionless variables defined as $r/a=r_0, \frac{a_{p/m}}{a}=a_0^{p/m}, m a=m_0$. The equations of motion can then be written as 
\beq
\phi_+(x_2+a_{p/m})\left(\frac{1}{2a_0^{p/m}}+\frac{r_0}{2(a_0^{p/m})^2}\right)&-&\phi_+(x_2-a_{p/m})\left(\frac{1}{2a_0^{p/m}}-\frac{r_0}{2(a_0^{p/m})^2}\right)\nonumber\\
&+&\left(m_0-\frac{r_0}{(a_0^{p/m})^2}+r_0\sum_{\mu=0,1}(\cos(p_{\mu}a)-1)\right)\phi_{+}(x_2)=0\nonumber\\
\label{eom1}
\eeq
for $x_2>0$ and $x_2<0$. I deliberately avoid writing the equation of motion for the `$-$' chirality mode as there are no normalizable solutions for it. There is one more equation which arises from the fact that the lattice spacing is discontinuous across $x_2=0$ and it is given by
\beq
\frac{\phi_+(a_0^p)}{(a_0^p+a_0^m)}\left(1+\frac{r_0}{a_0^p}\right)+\frac{\phi_+(-a_0^m)}{(a_0^p+a_0^m)}\left(-1+\frac{r_0}{a_0^m}\right)+m_0\phi_+(0)\nonumber\\
-r_0\,\frac{\phi_+(0)}{a_0^p a_0^m}+r_0\sum_{\mu=0,1}\left(\cos(p_{\mu}a)-1\right)\phi_+(0)=0.
\label{eom2}
\eeq
The E.O.M in Eq. \ref{eom1} are solved by an ansatz of the form 
\beq
\phi_+(x_2)&=&
\begin{cases}
z_{p}^{\frac{x_2}{a_p}}, & \text{for}\,\,\,\, x_2>0\\
z_{m}^{\frac{x_2}{a_m}} , & \text{for}\,\,\,\, x_2<0
\end{cases}\nonumber\\
\eeq
where $z_p$ and $z_m$ are given by
\beq
z_{p/m,\pm}&=&\frac{-\left(m_0-\frac{r_0}{(a_{0}^{p/m})^2}+r_0\sum_{\mu=0,1}(\cos(p_{\mu}a)-1)\right)}{\frac{1}{a_{0}^{p/m}}\left(1+\frac{r_0}{a_{0}^{p/m}}\right)}\nonumber\\
&\pm&\frac{\sqrt{\left(m_0-\frac{r_0}{(a_{0}^{p/m})^2}+r_0\sum_{\mu=0,1}(\cos(p_{\mu}a)-1)\right)^2+\frac{1}{(a_{0}^{p/m})^2}\left(1-\frac{r_0}{a_{0}^{p/m}}\right)\left(1+\frac{r_0}{a_{0}^{p/m}}\right)}}{\frac{1}{a_{0}^{p/m}}\left(1+\frac{r_0}{a_{0}^{p/m}}\right)}.\nonumber\\
\label{sol1}
\eeq
As seen from Eq. \ref{sol1} there are two independent solutions to Eq. \ref{eom1}. The most general solution for $\phi_+$ is then given by an arbitrary linear combination of the form 
\beq
\phi_+=\begin{cases}
A_+ z_{p,+}^{\frac{x_2}{a_p}}+ B_+ z_{p,-}^{\frac{x_2}{a_p}}, & \text{for}\,\,\,\, x_2\geq 0\\
A_- z_{m,+}^{\frac{x_2}{a_m}}+ B_- z_{m,-}^{\frac{x_2}{a_m}}, & \text{for}\,\,\,\, x_2\leq 0\\
\end{cases}
\eeq
where $A_+$ and $B_+$ are nonzero only when $|z_{p,+}|^{\frac{x_2}{a_p}}, |z_{p,-}|^{\frac{x_2}{a_p}}$ are normalizable for $x_2>0$ and $A_-$ and $B_-$ are nonzero only when $|z_{m,+}|^{\frac{x_2}{a_m}}, |z_{m,-}|^{\frac{x_2}{a_m}}$ are normalizable for $x_2<0$. However, $A_+, A_-, B_+$ and $B_-$ satisfy three more constraints. One of them is that $|\phi_+|^2$ integrated over $x_2$ normalizes to $1$. Also, since $\phi_+(0)$ is single valued, 
\beq
A_+ + B_+=A_- + B_-.
\label{cond2}
\eeq
The third condition is that $\phi_+$ has to satisfy Eq. \ref{eom2}. With this I can now proceed to analyze the number and chirality of the zero mode solutions on the wall.
In order for normalizable solutions to exist I need, 
\beq
&&|z_{p,\pm}|<1, \nonumber\\ 
&&|z_{m,\pm}|>1.
\label{ineq}
\eeq 
It is important to note that there are no solutions to the Eq. \ref{eom2} if at least two of the conditions in the inequality of \ref{ineq}  
are violated. Although, whether these conditions are met can be checked for any mode of interest I will concentrate on the modes centered around the Brillouin zone corners: $\{p_0 a=0, p_1 a=0 \}$, $\{p_0 a=0, p_1 a=\pi \}$, $\{p_0 a=\pi, p_1 a=0 \}$ and $\{p_0 a=\pi, p_1 a=\pi \}$. Also, I will only focus on $r_0=1$ and $m_0=1, 3, 5$. The reason behind restricting the analysis to these $m_0$ values is merely convenience as similar analysis can be performed for any other values of $m_0$. 
 In order to understand whether there exist zero mode solutions for any set of values for $a_0^p$ and $a_0^m$ it is useful to recognize that the functional form of $z_{p,\pm}$ and $z_{m,\pm}$ as a function of $\frac{1}{(a_{0}^{p})^2}$ and $\frac{1}{(a_{0}^{m})^2}$ are the same.  A convenient way to analyze the roots $z_{p/m,\pm}$ is then to define $z_{\pm}$ which has the same functional form in terms of the variable $\frac{1}{(a_0)^2}$ 
\beq
z_{\pm}\left(\frac{1}{(a_0)^2}\right)&=&\frac{-\left(m_0-\frac{r_0}{(a_{0})^2}+r_0\sum_{\mu=0,1}(\cos(p_{\mu}a)-1)\right)}{\frac{1}{a_{0}}\left(1+\frac{r_0}{a_{0}}\right)}\nonumber\\
&\pm&\frac{\sqrt{\left(m_0-\frac{r_0}{(a_{0})^2}+r_0\sum_{\mu=0,1}(\cos(p_{\mu}a)-1)\right)^2+\frac{1}{(a_{0})^2}\left(1-\frac{r_0}{a_{0}}\right)\left(1+\frac{r_0}{a_{0}}\right)}}{\frac{1}{a_{0}}\left(1+\frac{r_0}{a_{0}}\right)}.\nonumber\\
\label{sol12}
\eeq
In order to check whether there exists normalizable solutions for a particular mode one then has to analyze the behavior of $z_{\pm}$ as a function of $\frac{1}{a_0}$ i.e. the existence of a zero mode depends on whether there exists $a_0=a_0^p$ and $a_0=a_0^m$ for which one can find $|z_+(\frac{1}{(a_0^p)^2})|<1$, $|z_-(\frac{1}{(a_0^p)^2})|<1$, $|z_+(\frac{1}{(a_0^m)^2})|>1$, $|z_-(\frac{1}{(a_0^m)^2})|>1$ which satisfy the conditions in Eq. \ref{ineq} listed above.

Note that, in order for the bound in inequality \ref{ineq} to be obeyed, the curves $|z_+(\frac{1}{(a_0)^2})|$ or $|z_-(\frac{1}{(a_0)^2})|$ as a function $a_0$ has to intersect the line $|z_+|=1$ for a real and positive $a_0$. 
For $m_0=1$, none of the corners of the Brillouin zone except the mode $\{p_0 a=0, p_1 a=0\}$ exhibit a real positive $a_0$ at which $|z_+(\frac{1}{(a_0)^2})|=1$. This point at which $|z_+(\frac{1}{(a_0)^2})|=1$ is at $a_0=\sqrt{2}$. For $m_0=3$, for the mode $\{p_0 a=0, p_1 a=0\}$, $|z_+(\frac{1}{(a_0)^2})|=1$ at $a_0=\sqrt{\frac{2}{3}}$ and for the mode $\{p_0 a=\pi, p_1 a=0\}$ and $\{p_0 a=0, p_1 a=\pi\}$, $|z_+(\frac{1}{(a_0)^2})|=1$ at $a_0=\sqrt{2}$. However, for $\{p_0 a=\pi, p_1 a=\pi\}$ there is no $a_0$ for which $|z_{\pm}(\frac{1}{(a_0)^2})|=1$. For $m_0=5$ one can find real and positive $a_0$ satisfying $|z_{+}(\frac{1}{(a_0)^2})|=1$ for all the corners of the Brilluoin zone. For the mode $\{p_0 a=0, p_1 a=0\}$ this point is at $a_0=\sqrt{\frac{2}{5}}$. For the modes $\{p_0 a=0, p_1 a=\pi\}$ and $\{p_0 a=\pi, p_1 a=0\}$ this point is at $a_0=\sqrt{\frac{2}{3}}$. And for the mode $\{p_0 a=0, p_1 a=\pi\}$ this point is at $a_0=\sqrt{2}$.

This implies that for $m_0=1$, for certain values of $a_p$ and $a_m$ one can find normalizable solution for the mode $\{p_0 a=0, p_1 a=0\}$. Similarly, for $m_0=3$, it is possible to obtain normalizable solutions for the modes $\{p_0 a=0, p_1 a=0\}$, $\{p_0 a=\pi, p_1 a=0\}$ and $\{p_0 a=0, p_1 a=\pi\}$. For $m_0=5$, all the BZ corner modes can have normalizable solutions for appropriately chosen values for $a_p$ and $a_m$.

In what follows I discuss a few choices of $a_0^p$ and $a_0^m$ for $m_0=1, 3$ and $5$ and demonstrate that the number and net chirality of 
normalizable zero mode solutions is consistent with the C.S. levels obtained in subsection \ref{sub2} in Fig. \ref{fig1} as a function of lattice anisotropy for various values of $m_0$.
\begin{figure}[h!]
\centering
\includegraphics[width=\textwidth]{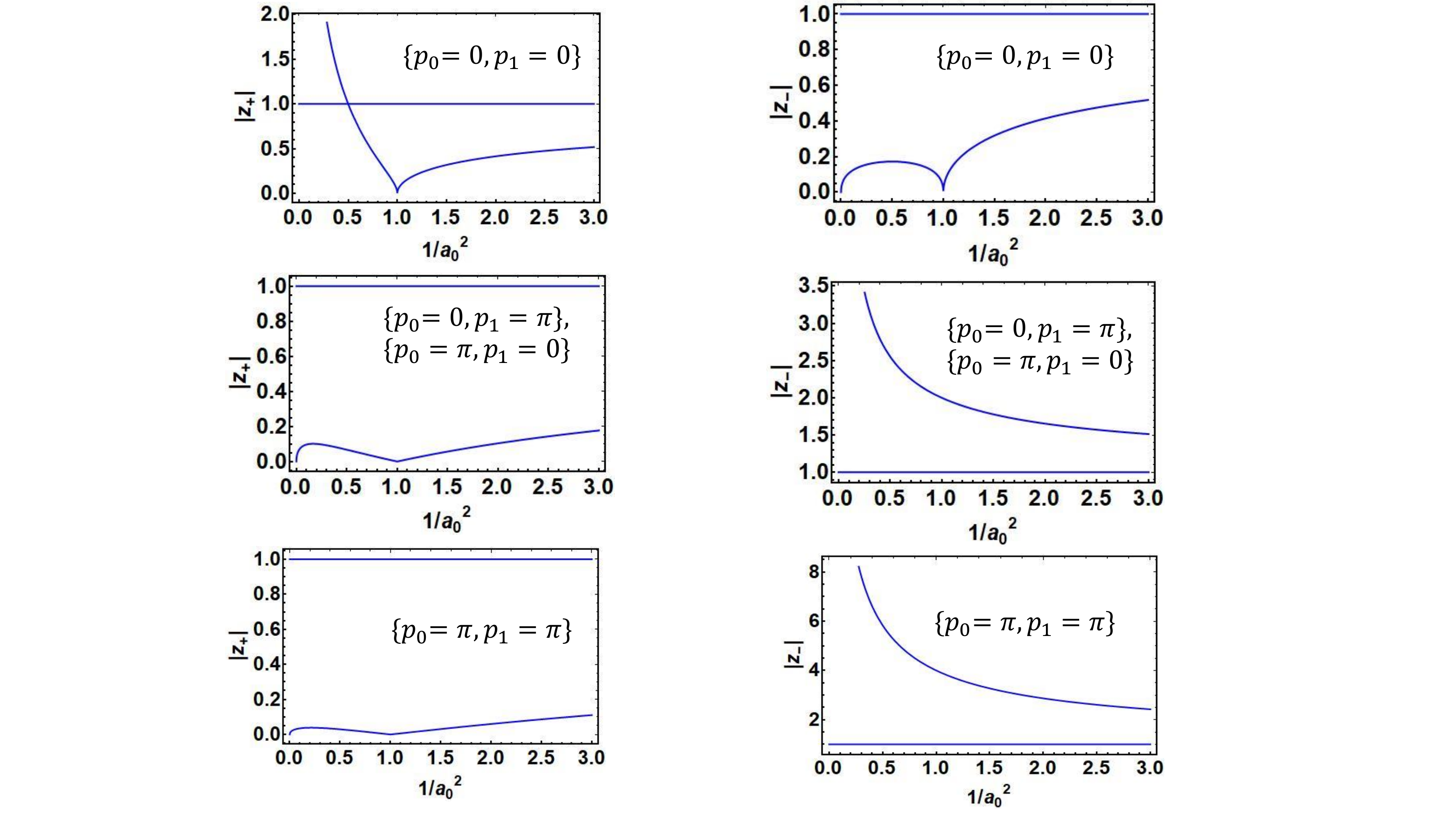}
\caption{$|z_{\pm}|$ as a function of lattice spacing for $m_0=1$.}
\label{fig3a}
\end{figure}
First consider $m_0=1$. As can be seen from Fig. \ref{fig1}, for $m_0=1$, with $(1/a_0^m)^2 <1/2$ and $(1/a_0^p)^2 > 1/2$, there is a C.S. current of level $+1$ flowing  to the wall from the $x_2>0$ bulk and There is no current in the $x_2<0$ bulk. This suggests that there must exist net chirality of $+1$ on the wall. As is clear from Fig. 3, this indeed is the case. There are no normalizable solutions satisfying Eq. \ref{eom2} with any choice of $a_0^{p/m}$ for  $\{p_0 a=\pi, p_1 a=0\}$, $\{p_0 a=0, p_1 a=\pi\}$ and $\{p_0 a=\pi, p_1 a=\pi\}$. However, the mode $\{p_0 a=0, p_1 a=0\}$ is normalizable on the wall so long as $(1/a_0^m)^2 <1/2$ and $(1/a_0^p)^2 > 1/2$. As a result, anomaly inflow works out perfectly. 

\begin{figure}[h!]
\centering
\includegraphics[width=\textwidth]{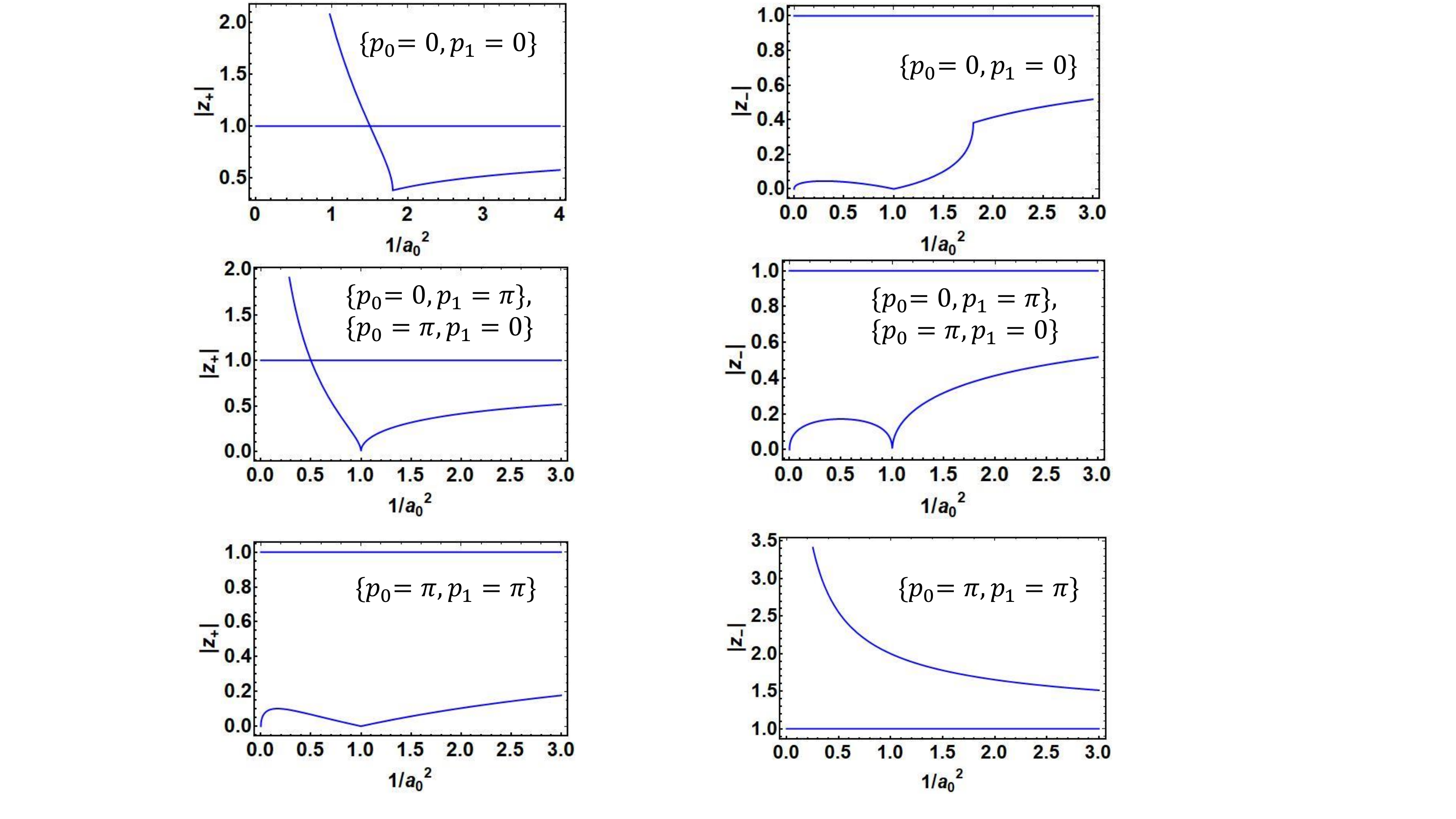}
\caption{$|z_{\pm}|$ as a function of lattice spacing for $m_0=3$.}
\label{fig3b}
\end{figure}

For $m_0=3$, something a bit more interesting takes place. For example, for $(1/a_0^m)^2 <1/2$ and $3/2>(1/a_0^p)^2 > 1/2$ there is a C.S. current of level $2$ flowing away from the wall for $x_2>0$ whereas there is no C.S. current for $x_2<0$. This suggests that the wall must harbor net chirality of $-2$ so as to maintain current conservation. From Fig. \ref{fig3b}, it is clear that for $(1/a_0^m)^2 <1/2$ and $3/2>(1/a_0^p)^2 > 1/2$, neither $\{p_0 a=0, p_1 a=0\}$ or $\{p_0 a=\pi,p_1 a=\pi\}$ are normalizable. However, $\{p_0 a=0, p_1 a=\pi\}$ and $\{p_0 a=\pi, p_1 a=0\}$ are, which is consistent with current conservation or anomaly inflow. When $(1/a_0^p)^2 > 3/2$ and $1/2<(1/a_0^m)^2 < 3/2$, there is a C.S. current of C.S. level $2$ flowing to the wall in $x_2<0$ whereas there is a C.S. current of level $1$ flowing away from the wall in $x_2>0$. For anomaly inflow to work, the net chirality of zero modes on the wall be $+1$. As can be seen from Fig 3. for the choice of $(1/a_0^p)^2 > 3/2$ and $1/2 <(1/a_0^m)^2 < 3/2$, $\{p_0 a=0, p_1 a=\pi\}$, $\{p_0 a=\pi, p_1 a=0\}$ and $\{p_0 a=\pi, p_1 a=\pi\}$ are not normalizable where as $\{p_0 a=0, p_1 a=0\}$ is. As a result the net chirality on the wall indeed is $+1$.  

If on the other hand, $(1/a_m)^2 <1/2$ and $(1/a_p)^2 > 3/2$, there is a net C.S. current of level $1$ is flowing away from the wall from $x_2>0$ and there is no current for $x_2<0$. This requires the wall to harbor zero modes such that the net chirality on the wall is $-1$. It is easy to see from Fig. \ref{fig3b} that for this choice of $a_0^p$ and $a_0^m$, $\{p_0 a=0, p_1 a=\pi\}$, $\{p_0 a=\pi, p_1 a=0\}$ and $\{p_0 a=0, p_1 a=0\}$ are normalizable, but $\{p_0 a=\pi, p_1 a=\pi\}$ is not. Indeed the net chirality of the modes on the wall add up to $-1$. Note that the number of zero modes and the net chirality don't coincide. Hence, the number of topologically protected zero mode is $1$ and it is a linear combination of the modes $\{p_0 a=0, p_1 a=\pi\}$ and $\{p_0 a=\pi, p_1 a=0\}$.

\begin{figure}[h!]
\centering
\includegraphics[width=\textwidth]{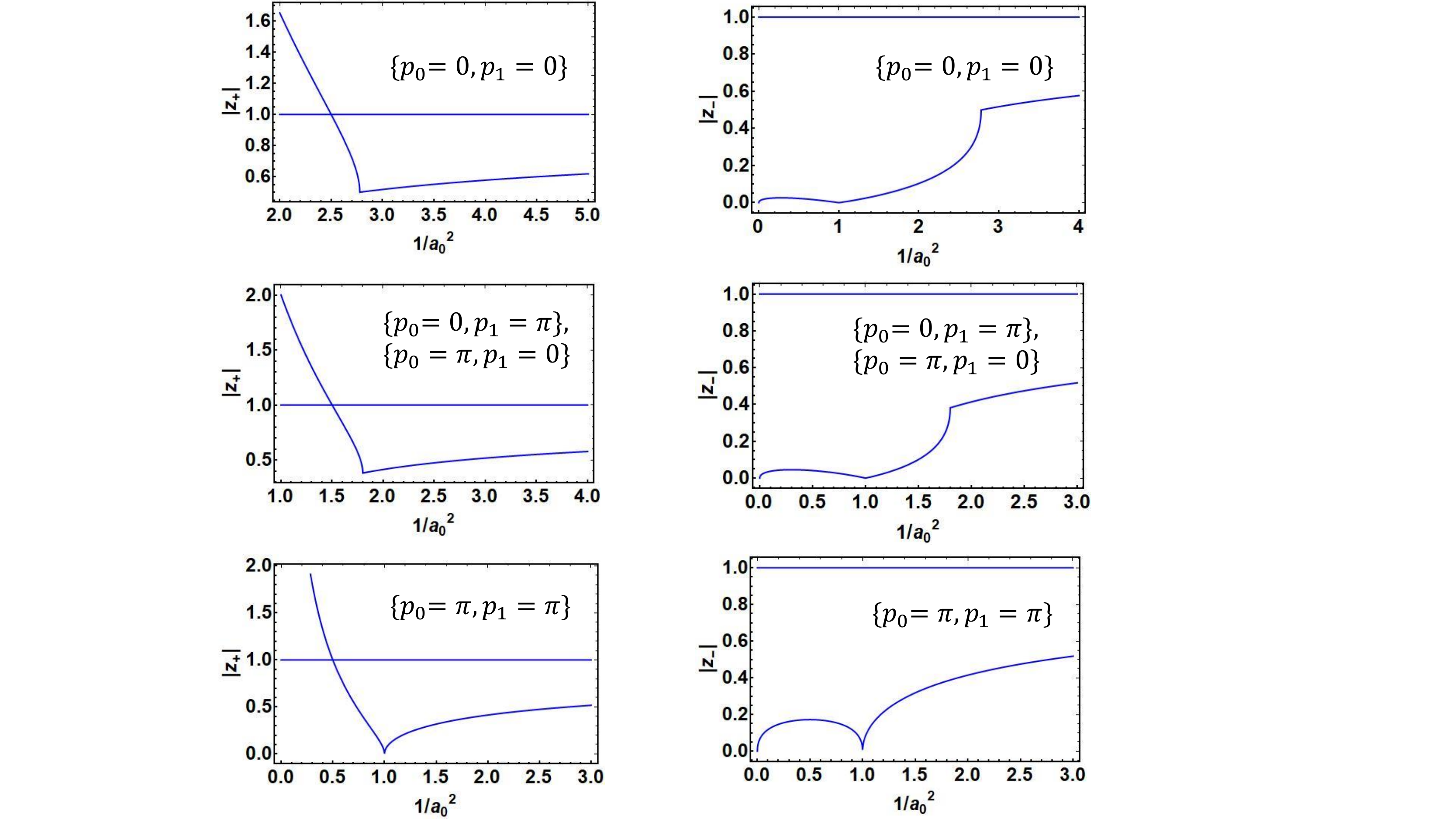}
\caption{$|z_{\pm}|$ as a function of lattice spacing for $m_0=5$.}
\label{fig3c}
\end{figure}

In the case of $m_0=5$, there are again quite a few interesting possibilities as can be seen from Fig. \ref{fig1} and some of these possibilities I discuss below. 
When $\frac{1}{(a_0^m)^2}<\frac{1}{2}$ and $\frac{1}{2}<\frac{1}{(a_0^p)^2}<\frac{3}{2}$ there is no current in $x_2<0$ whereas there is a C.S. current of level $1$ flowing to the wall in $x_2>0$ bulk, thus requiring the net chirality on the wall to be $1$. From Fig. \ref{fig3c}, for these values of $a_0^p$ and $a_0^m$, $\{p_0 a=0,p_1 a=0\}$, $\{p_0 a=0,p_1 a=\pi\}$, $\{p_0 a=\pi,p_1 a=0\}$ are not normalizable whereas $\{p_0 a=\pi,p_1 a=\pi\}$ is. This is in agreement with anomaly inflow. For $\frac{1}{2}<\frac{1}{(a_0^m)^2}<\frac{3}{2}$ and $\frac{3}{2}<\frac{1}{(a_0^p)^2}<\frac{5}{2}$, there is a C.S. current of C.S. level $1$ flowing away from the wall from both sides of the wall. This implies that the wall should have a net chirality of $-2$. From Fig. \ref{fig3c}, one can see this indeed is the case since for these values of $a_0^{p}$ and $a_0^m$ only $\{p_0 a=0, p_1 a=\pi\}$ and $\{p_0 a=\pi, p_1 a=0\}$ are normalizable whereas $\{p_0 a=0,p_1 a=0\}$ and $\{p_0 a=\pi, p_1 a=\pi\}$ are not.  
If $\frac{3}{2}<\frac{1}{(a_0^m)^2}<\frac{5}{2}$ and $\frac{5}{2}<\frac{1}{(a_0^p)^2}$, there is a net C.S. current of level $+1$ flowing to the wall for $x_2<0$ and none in $x_2>0$, thus requiring a net chirality of $+1$ on the wall. From Fig. \ref{fig3c}, it can be seen that for $\frac{3}{2}<\frac{1}{(a_0^m)^2}<\frac{5}{2}$ and $\frac{5}{2}<\frac{1}{(a_0^p)^2}$, $\{p_0 a=0, p_1 a=0\}$ is normalizable whereas $\{p_0 a=\pi, p_1 a=\pi\}$,  $\{p_0 a=0, p_1 a=\pi\}$, $\{p_0 a=\pi, p_1 a=0\}$  aren't. This is again consistent with anomaly inflow requirements. All other possibilities of various $a_0^p$ and $a_0^m$ which I do not discuss here satisfy anomaly constraints as well. 
\section{Chern insulator} The similarity between the physics of topological phases in Chern insulators and the domain wall fermion (DWF) construction of lattice gauge theory for cubic lattices becomes clear 
if one writes down the Hamiltonian for the two-band Chern insulator \cite{PhysRevB.78.195424}
\beq
H(\mathbf{k})=\sin(k_x)\sigma_x + \sin(k_y)\sigma_y +(\mathcal{M}+2+(\cos(k_x)-1)+(\cos(k_y)-1))\sigma_z.
\label{CI}
\eeq 
Note that, this Hamiltonian involves momenta that have been scaled by the lattice spacings for a square lattice. In order to recognize the parallels between the lattice construction of DWFs and Chern insulators as described by Eq. \ref{CI}, one can set $a_{\mu}=1$ and $r=1$ in the inverse fermion propagator in Eq. \ref{Dfermi} and it is then clear that the Hamiltonians of the two systems coincide with the identification of $m=\mathcal{M}+2$. Although the Chern insulator and lattice DWF construction are analogous, one distinguishing feature between the two is that the time direction is discretized in the latter and is not so in the former. The Chern number of the Hamiltonian in Eq. \ref{CI} is found to toggle between the following values as a function of the parameter $\mathcal{M}$
\beq
C=\begin{cases} 1, \,\,\,\,\,\,\,\,\,\,\,\,\,\,\,0<\mathcal{M}<2\\
-1, \,\,\,\,\,\,-2<\mathcal{M}<0\\
0, \,\,\,\,\,\,\,\,\,\,\,\,\,\,\,\text{otherwise.}
\end{cases}
\label{Chern}
\eeq 
The discrete changes in the Chern number correspond to topological transitions as a function of the parameter $\mathcal{M}$. These transitions are analogous to the topological transitions observed in the lattice construction of DWFs using a Wilson fermion of mass $m$ where the Wilson parameter is set to be equal to the lattice spacing. Note that the Chern-Simons levels that the DWF theory alternates between i.e. $1, -2$ and $1$ are different from Chern numbers observed in Eq. \ref{Chern}. This difference can be attributed to the time direction being discretized in the DWF context.

It is now clear that the analysis of the rectangular lattices and the corresponding edge modes as elaborated in the previous sections can be replicated for the Chern insulators too. Again, the only difference between the two analysis is going to stem from the time direction being discrete in the lattice gauge theory and it being continuous in Chern insulators.  

Note that, there have been some previous work on lattice models with anisotropic hopping parameter in the context of second order topological insulators as discussed in \cite{PhysRevB.100.115403}. Interestingly, these models and the lattice constructions in this paper share some common features. One of these common features is that in the presence of an anisotropy for the hopping parameter, a `non-topological' phase of Chern number zero emerges between the two topological phases with Chern numbers $+1$ and $-1$. This is similar to what is seen in Fig. \ref{atv1} for $\tilde{a}=0.8$ and $\tilde{r}=1$. Similarly, another common feature is the existence of zero mode solutions to the equations of motion that are not topologically protected. However, one of the main differences between the construction presented in this paper and in \cite{PhysRevB.100.115403} is that \cite{PhysRevB.100.115403} employs square lattices as opposed to rectangular lattices considered in this paper. 
Another difference lies in the continuum limit of the relevant models in \cite{PhysRevB.100.115403} where the hopping parameter breaks rotational invariance in the continuum. 

\section{Conclusion and future work} The analysis of Wilson fermions on a rectangular lattice in this paper reveals new topological phases that are not accessible on a cubic lattice. One of the most interesting features of the analysis is associated with the edge modes where I find that the number of topologically protected edge modes does not always equal the number of zero mode solutions to the Dirac equation, i.e. the equations of motion can admit zero mode solutions of opposite chirality on a $1+1$ dimensional discontinuity. Such a discontinuity can be a domain wall in the fermion mass or it can be a lattice defect with an abrupt change in lattice spacing in the direction perpendicular to the wall. Another remarkable feature of the construction on rectangular lattice is that lattice defects across which lattice spacing changes abruptly can exhibit chiral zero mode solutions in the absence of any domain wall in the fermion mass. The distinct features of rectangular lattices noted in this paper maybe interesting to explore in lattice simulations which have so far dealt with cubic lattices while exploring topological phases of Wilson fermions. Simulating the domain wall in fermion mass and the lattice spacing defect considered in this paper will require the transverse dimension to be finite \cite{Shamir:1993zy} in which case it may be interesting to consider if and how the finite extent of the transverse direction interplays with the lattice anisotropy in dictating the phase diagram. 

Note that, the calculations in this paper are performed at weak coupling which corresponds to the continuum limit of the lattice. The fate of the various phases obtained in this paper away from the weak coupling limit can be explored in lattice simulations and could be interesting avenue for future work. Another important consideration here is the gauge sector of the theory. Although this paper focuses on analyzing Wilson fermion coupled to a $U(1)$ gauge theory, the analysis can very well apply to $SU(N)$ gauge theories. It will therefore be interesting to explore how the phase diagram behaves as a function of the gauge coupling and lattice anisotropy for QCD (Quantum Chromodynamics) and QCD like theories in $2+1$ and $4+1$ dimensions.

It will also be instructive to explicitly compute the profile of the edge mode wave functions for the lattice defect across which the lattice spacing transverse to the defect changes abruptly. The rates of the exponential fall off of the wave function away from this defect are in general unequal on the two sides of the defect. It is conceivable that the Chern-simons levels on the two sides of the lattice defect are in some way correlated with the exponential fall off of the wave functions.  

Another possible direction for a follow up project involves working out the Chern insulating transitions on a rectangular spatial lattice without discretizing the time dimension. This exercise will demonstrate how a rectangular spatial lattice responds to a changing lattice anisotropy. Similarly, the two types defects considered in this paper are worth exploring in the presence of a spatially rectangular lattice with a continuous time dimesnion.

\section{Acknowledgement}
I thank Thomas Iadecola and Michael Wagman for useful comments.

\bibliographystyle{utphys}
\bibliography{anisotropic}

 \end{document}